\documentclass[12pt]{article}
\newcommand{\be}{\begin{equation}}
\newcommand{\ee}{\end{equation}}

\begin{document}
\baselineskip18pt

\title{Impenetrable Barriers and Canonical Quantization}

\author{Piotr Garbaczewski\thanks{Email: pgar@proton.if.wsp.zgora.pl}\\
Institute of  Physics, Pedagogical University \\ PL-65 069 Zielona
G\'{o}ra,  Poland \\
and\\
Witold Karwowski \\
Institute of Theoretical Physics, University of Wroc{\l}aw, \\
PL-50 205 Wroc{\l}aw, Poland}
\maketitle

\begin{abstract}
We address a conceptual issue of reconciling the  traditional canonical quantization
framework  of quantum theory  with the spatially restricted  quantum dynamics
and the related spectral problems for confined and global observables of the
quantum system.
\end{abstract}

 \section{Motivation}

Modern technologies enable one to enslave  individual quantum  particles in various traps
for long time intervals. Manipulations with nanostructures involve  a fine tuned control
 of   quantum wells  shape  and depth that has  a decisive influence on
 blocking or enabling various transport (in fact,  tunelling) phenomena.
In all those cases a fairly pragmatic usage of the traditional quantum mechanical formalism
 shows an acceptable explanatory/predictive power.

A proliferation of papers on various aspects of the infinite
potential well, \cite{stroud} - \cite{klauder}, and  on
sophisticated "exercises in exact quantization" on half-line,
\cite{voros} motivates our renewed interest in reconciling the
canonical quantization principles with the \it sole \rm analysis
of  well posed spectral problems for the Hamilton  operator,
with   Dirichlet or  Neumann boundary data.
 The latter,  purely  spectral (spectroscopy  oriented) attitude is
quite strongly represented in the modern literature pertaining to
 mesoscopic  systems, \cite{stock,hurt,chavel}.
In the study of so-called quantum billiards
(integrable, pseudointegrable and/or chaotic) and of the related
microwave cavities, one investigates  eigenvalue problems for the Laplacian on
a connected and compact  domain of arbitrary shape in $R^2$, in particular
with Dirichlet boundary condition.
Normally, that analysis is  devoid of any  "spurious"  canonical
quantization input and focuses on statistical properties of eigenvalue series,
predominantly with emphasis on the semiclassical regime.

 A major surprise in this context is that   a careful  analysis of the involved
\it conceptual \rm background  reveals  apparent inconsistencies and paradoxes
\cite{valent,gori,klauder}, if one seriously attempts to reconcile mathematical
models of  trapping (space-time localization) with the   apparatus of
\it canonical quantization, \rm that is commonly believed to underlie  the
traditional quantum mechanical framework.

The main  objective  of  the present paper is  an analysis  of the
\it restricted particle dynamics \rm
in quantum theory and its relation to the canonical quantization   (carried out
in the standard Schr\"{o}dinger representation).
For clarity of presentation most of our discussion will be confined to quantum
mechanics on the real line with a reference  Hilbert space $L^2(R^1)$, although
much of the argumentation can be directly adopted to higher dimensions.

For a quantum particle that can be anywhere on the real line, there are a  priori
 no restrictions on wave functions $\psi $, of any external origin, that would
keep a particle confined  within certain interval on $R^1$  - for a finite time or
indefinitely.
We are interested in the situation when the quantum particle is so restricted
that it cannot be  situated on certain parts of the real line  \it at any time,  \rm or
in the least there is  \it no \rm communication (tunneling  \cite{karw} or any
other conceivable form of quantum mechanical transport) between  those
 parts and their complement on $R^1$.

Typical examples of such circumstances
 in quantum theory are provided by introducing impenetrable walls (which can be
consistently  interpreted as an idealisation  of the trapping enclosures on $R^1$).
Such barriers are externally imposed  and need to have an effect on the
physical characteristics of the  quantum system which are conventionally associated with
the notion of the (pure or mixed) state of the system and  relevant observables .
 Less spectacular but important examples of impenetrability are related to the existence of
 nodes  or nodal curves (surfaces) of wave functions  (cf. the stationary state issue),
\cite{karw,olk}.

Wave functions (solutions of the Schr\"{o}dinger equation adopted to a chosen situation)
carry a probabilistic information about the space-time localization of a particle.
However, this very (localization) notion  comes from first assuming that there are
the primitive (primordial) kinematic  observables
related to position and momentum (selfadjoint  position and momentum
operators with a continuous spectrum) which are inseparable from the concept of
\it canonical quantization. \rm  It is the \it emergent \rm (secondary)
energy observable that  sets the unitary  (Schr\"{o}dinger) dynamics  for
the quantum problem,  where $\psi (x,t)$ ultimately appears as a solution of
the differential equation with suitable  initial/boundary data.
Precisely at  this   point  an evident clash occurs between
quantum mechanical pragmatism and  the deep need for an overall consistency of the
formalism employed.

Traditionally one expects that the algebra of observables for a quantum mechanical
problem contains  suitable self-adjoint operators and the generator of unitary
dynamics - the Hamiltonian,  needs to be among them. The self-adjointness property is
required because of the  spectral theorem which sets a unique link between an
operator and its family of spectral projections. That in turn allows
to state unambigous  "elementary questions" about properties of a physical system
(by invoking projection operators to ask for  a probability of  locating
 a particle in a  given interval, to find its momentum within certain range etc.)

An apparent problem  can be seen at once,  if we consider a particle on $R^1$
 that is e.g.
permanently residing between two  impenetrable barriers  (rigid walls), set at  points
$\pm b,\, b\in R^+$. Clearly, that enforces a condition that $\psi (x,t) =0$ for all
$|x|\geq b$.
One may think that a Hamiltonian can be simply defined "as it is ", like e.g.
a differential operator $-{\hbar ^2\over {2m}}{d^2\over {dx^2}}$,
 and then both in-between and outside of the impenetrable walls.
The point is that such a globally
 defined Hamiltonian is not a selfadjoint operator, \cite{schechter}.

A consistent introduction of the \it unitary \rm quantum dynamics needs  a careful
examination of self-adjoint extensions of otherwise merely  symmetric   operators
 and if there are many of them, we encounter a number of  \it inequivalent \rm
physical evolution problems associated with a unique for all cases
symmetric operator.

 Another obvious clash with the pedestrian intuition can be immediately invoked
if we  attach the name of "momentum operator" to the differential expression
 ($-i\hbar {d\over {dx}}$)  which has a continuous spectrum
 in reference to quantum particle on $R^1$. Other representatives of this operator, but
with  discrete spectra,  would appear if to follow a
 typical "particle in the box" procedure with periodic boundary conditions.

There is yet  another clash involved: "momentum operator eigenfunctions  do not exist
in a box with rigid (!) walls, since then they would vanish everywhere", \cite{schiff}.
And still another clash occurs, within the fully-fledged pragmatism of the
grand text-book  discussion, \cite{cohen}, where the "momentum measurement" and the
distribution
of (continuous !) momentum values in stationary state  of a particle in an
infinite potential well is
considered in minute detail  as a re-examination of the subject  "from a physical
point of view".
In fact, in Ref. \cite{cohen}  an explicit answer is formulated for what is "the probability
of a measurement of the momentum $P$ (observable) of the particle yielding a result between
$p$ and $p+dp$". That involves an explicit usage of the Fourier integral for
spatially confined wave packets and clearly derives from assuming that the proper
meaning of the momentum operator is that it generates spatial translations and the
related unitary group of transformations whose arena is the whole of $L^2(R^1)$.

Just an opposite extreme for the infinite well problem was  verbalized in the very
recent Ref. \cite{klauder}: "Next we turn to the momentum representation.
Since the spectrum of
the operator $P$ is discrete, the Hilbert space in the momentum representation
reduces  to the space $l^2$ of square summable sequences.
This is just a reformulation of the theory of Fourier
series as opposed to the Fourier integral that makes a transition between the
position and momentum  representation  for a quantum mechanics on the full
line $R^1$ (nothing of course, forbids one to
take the Fourier integral transform of the infinite well wave
function $\psi (x,t)$, but the
result is just  a mathematically equivalent version of the same object, \it not \rm
the momentum representation wave function)".
The trouble emerges from the fact that  the self-adjoint Hamiltonian operator with the
rigid wall (vanishing) Dirichlet  boundary data does not at all coincide with an
operator ${P^2\over {2m}}$, where $P$ is  taken  as a selfadjoint momentum operator.
The latter observable respects
the periodic boundary data and as such gives rise to a quantization \it on a circle, \rm
with ${P^2\over {2m}}$ representing the so-called plane rotator  Hamiltonian, and thus
bears no trace of  reference to the original infinite well problem.

While  on $R^1$, we must also address an issue of the exterior of  impenetrable barriers
 set for the infinite well problem, or a simpler  case of one permanently installed
 impenetrable barrier  dividing $R^1$ into two non-communicating  segments.
 Then a quantum  particle, if at all in existence, is  restricted to
stay effectively on the half-line, \it either \rm positive or negative.
In that case, the positive and
negative semiaxis correspond to disjoint, completely independent
quantum mechanical problems. The Schr\"{o}dinger particle on a
half-line is a generic case, \cite{simon}, where we have nicely
elucidated major obstacles that hamper the "pragmatic" usage of
symmetric operators, without paying  attention to their
self-adjointness,   and in particular to the very existence of
the unitarily implemented dynamics.

In turn, that example is diagnostic to the
proper handling of quantum phenomenology, when the priority of  functional analysis
(operator self-adjointenss issue) is put forward against physical  intuition.
Clearly, the problem of self-adjoint extensions of both the free $-{{\hbar ^2}\over {2m}}
{{\partial ^2}\over {\partial x^2}}$ and perturbed (conservative) Hamiltonian on
the half-line  is a classic.
We would like to recall   at this point that  a symmetric momentum
operator $-i\hbar {\partial \over {\partial x}}$  not only is \it not \rm self-adjoint,
but even has \it no \rm self-adjoint extensions when analyzed exclusively in
$L^2(R^+)$ or $L^2(R^-)$. That made the authors of Ref. \cite{valent} to conclude:
"the momentum is not a measurable quantity in that situation !".

Leaving aside a delicate issue of what is actually   meant by the "momentum measurement"
in the half-line or specific mesoscopic (trapping) contexts,
 we take   the view that the major conceptual obstacle  behind the previous
statement (and similar  "paradoxes")  involves an improper handling of kinematic
observables.

\section{Barriers}

A quantum particle that is  trapped inside the infinite well
$0\leq x\leq \pi $ must have its wave function equal to zero
outside the well. That is usually enforced by assuming that the
potential  $V(x) = \infty $ on the  complement of an open
interval $(0,\pi )$ in $R^1$ hence for $x\geq \pi $ and $x\leq
0$,  while $V(x) = 0$ between the impenetrable barriers. In view
of an infinite discontinuity of the potential, the wave function
$\psi (x)$ must vanish  for   $x\geq \pi $ and $x\leq 0$ and no
restrictions are imposed on its gradient at the interval
boundaries. Effectively, as far as  the sole trapping is concerned,
one  rather ignores \it any restrictions \rm that would
extend to the rest of the real line (although definitely one
 should \it not \rm ignore that the "rest" of the  space
 itself is in existence) and considers
the sufficient conditions for permanent trapping: $\psi (0)= \psi
(\pi ) = 0$. Periodized well boundary conditions $\psi (n\pi )=
0,\,   n\geq 1$ would be then conceivable as a simplified model
of multiple traps and the whole exterior of a given trap would
still matter as an ingredient of the  formalism.

Even if  the particle is a priori confined in one concrete  trap,
we do   not accept the view that one may literally shrink the whole "quantum
world" from $R^1$ to the interior of that  concrete  interval and
thus to the Hilbert space $L^2([0,\pi])$ only. Although  all of
the pertinent quantum dynamics is  confined  to  the trap $(0,\pi
)$, cf. \cite{stroud}-\cite{klauder} and \cite{cohen}.

Our point of view is  supported by the following reasoning.
Given a normalized function $f\in L^2(R^1,dx)$, one may
consistently ask text-book  questions like "what is a probability
to observe a particle in the interval $M \subset R^1$" or  "what
is a probability to obtain the  result of momentum measurement in
 $K\subset R^1$". The answers are  standard (as far as we maintain
 a straightforward $R^1$ lore):
 \begin{equation}
P_{x\in M} =\int_M |f(x)|^2 dx\, \, , \,  \,  P_{p\in K} = (2\pi )
\int_K \tilde{f}(p)|^2 dp
\end{equation}
where $\tilde{f}$ is  the Fourier transform of $f$.

Let us specify
suitable domain  restrictions  for   position
and  momentum operators: $D(Q) = [f\in L^2(R^1); \int_{R^1}
|xf(x)|^2 dx < \infty ]$ and  $D(P) = [f\in L^2(R^1); \int_{R^1}
|p\tilde{f}(p)|^2 dp < \infty ]$.
Clearly, for $f\in D(QP)\bigcap D(QP)$ we can evaluate various
expectation values, and in particular  deduce  the Heisenberg
indeterminacy relation (in a pure state represented by $f$)
$\triangle P \triangle Q \geq {1\over 2}$ (up to the Plack
constant $\hbar $) which is a direct consequence of the canonical
quantization ansatz $[Q,P] \subset iI$.

If we consider the Heisenberg
inequality as one of conceptual cornerstones of quantum theory, there is no
 way to attribute a pure point spectrum to the physically interpretable
momentum operator nor admit bounded "position operators" as physically relevant objects.

In fact, a localization in  the interval,  which is trivially accomplished by invoking
spectral projections for   $Q$ whose continuous spectrum  extends through  $R^1$,
happens to be misinterpreted as  the need to define the whole of the quantum problem
to be confined to  that finite trap. Clearly, $Q$ when confined to the interval is
a bounded operator, but it is not a regular position operator but rather its localized
 spectral projection on the interval: $P_{[a,b]} Q P_{[a,b]}$.

All the above position-momentum issues  are purely kinematical and
thus  completely divorced from any assumptions about quantum
dynamics and the specific choice of the Hamiltonian.
\\

{\bf Example 1:}  We begin from  most traditional infinite well problem.
In that case, we assume a localization in the interval $[0,\pi ]$. Although the
boundary (Dirichlet) data demand  $\psi (0) = 0 = \psi (\pi)$, we  interpret them as
$\psi (x) = 0 $ for $x \leq 0$ and $x\geq \pi $. That clearly identifies
a specific localization on  $R^1$, instead  of an isolated "quantization on an interval"
issue.
From that point of view, a kinematic question  about a momentum information encoded
in wave functions with  those properties  automatically involves the  Fourier formalism of
 Eq. (1)  which perfectly works for spatially localized wave packets.
In particular, there does not make much sense to ask for the spectral resolution
 of the "momentum operator" $-i\hbar {d\over {dx}}$  restricted to  that spatial trap,
 since any  nontrivial function with the above Dirichlet boundary data is a
localized $L^2(R^1)$ wave packet and \it not \rm a plane wave.
When confined to the trap, the "free particle" Hamiltonian
operator eigenfunctions $\psi _n(x) = \sqrt{2\over \pi }
sin(n+1)x$
 are wave packets  and correspond to
$E_n= (n+1)^2 {\hbar ^2 \over {2m}}$, where  $n=0,1,...$.
(Notice  that the \it periodic  \rm boundary conditions would  not at all produce
a genuine solution of the momentum  operator eigenvalue
problem but rather the so-called quasi-momenta and corresponding quasi-momentum
eigenfunctions, of the form ${1\over \sqrt{\pi }}exp(i n \hbar x)$, with $x\in [0,\pi ]$
 which are $L^2(R^1)$ wave packets again and as such are amenable to  fully
fledged Fourier analysis (1).)\\

While passing to the problem of the  time-evolution we immediately
find that \it only \rm in the case of free motion  there is a
direct connection via spectral theorem between the momentum
and  Hamilton operators (like e.g. ${P^2/{2m}}$).
In other cases
there appear potentials and/or boundary conditions (eventually -- constraints
when dynamics on various manifolds is concerned).
Notice that the   boundary restrictions, sometimes can be interpreted as
related to null-set potentials,
\cite{karw}, and possibly as an interaction with the null-set dynamical systems.

Whatever the Hamiltonian may be, we can safely assume that  it is
bounded from below and that the value $0$ is the lowest point in
its  spectrum.

Given a Hamiltonian operator $H$, in view of the previously
mentioned permanent trapping problems, let us  consider the
following question: is there an open set $G\subset R^1$ such that
whenever $f\in D(H)$ then also   $\chi _Gf \in D(H)$, where $\chi
_G$ is the characteristic (indicator) function of $G$ ?

Assume tentatively that  the answer is positive. Then,   clearly
$\chi _G$ commutes with spectral projectors of $H$  and hence
with the unitary operator $exp(iHt)$. That would imply  an
invariance of the subspace $[f\in L^2(R^1); supp \, f \subset G]$
with respect to the time evolution.
The above localization issue can be re-told otherwise. Namely, in
that case the Hilbert space $L^2(R^1)$ and the operator $H$ split
into direct sums $L^2(R^1) = L^2(R\backslash G) \bigoplus L^2(G)$
and $H= H_1 \bigoplus H_2$, where  $H_1$ is selfadjoint in
$L^2(R^1\backslash G)$ and $H_2$ is selfadjoint in $L^2(G)$.

The physical and mathematical mechanisms leading to such reduction
of the dynamics can be illustrated by a number of  examples.\\

First,  we can supplement the previous Example 1 by defining
$H= - {{d^2}\over {dx^2}}$ through its specific
 domain $D(H) = [f\in AC^2(R^1); f, f', f'' \in L^2(R^1),
f(0) = f(\pi )]$.   The $AC^2$ notation refers to the absolute
continuity of the first derivative which guarantees  the
existence of the second derivative (in the sense of
distributions, as a measurable function). The  operator $\{ H,
D(H)\} $ is selfadjoint  and the decomposition $L^2(R^1) =
L^2(R\backslash G) \bigoplus L^2(G)$, together with
 $H= H_1 \bigoplus H_2$, holds true  for $G=[0,\pi ]$. Thus the traditional infinite well
problem  is  nothing else than the analysis  of $H_2$ in the space $L^2([0,\pi ])$.\\

{\bf Example 2:} Let us consider an   operator belonging to the
family of singular problems with the  centrifugal  potential
(possibly modified by the harmonic attraction),
\cite{calogero,olk}:
\begin{equation}
H = - {d^2 \over {dx^2}} + {1\over {[n(n-1)x^2]}}
\end{equation}
with $n\geq 2$ and $D(H) = [f=\in AC^2(R^1); f, f', f" \in
L^2(R^1), f(0)=0]$.  The above operator $H$ is known to be
self-adjoint. The projection operator   $P_+$ defined by $(P_+
f)(x)= \chi_{R^+}(x) f(x)$ clearly commutes  with H. The
singularity of the potential is sufficiently severe to \it
enforce \rm the boundary condition $f(0) =0$ (the \it generalized
\rm ground state function  (cf. Ref. \cite{berezanski})
 may be chosen for this scattering problem in the form
$\phi (x)= x^n$).\\
 Notice that we can here equivalently tell
about two \it separate \rm quantum problems, respectively  on
$R^+$ and $R^-$ (technically that refers to the degenerate ground state).
We deal here with  the most conspicuous illustration of
the fact that a particle cannot be "simultaneously" present in (shared by)
disjoint  trapping areas, e. g. cannot "live" on both sides of an impenetrable
barrier. Once trapped, a particle is
enslaved  in one    particular enclosure only and then cannot be detected in another.\\

{\bf Example 3:}
The classic Calogero-type problem is defined by
\begin{equation}
H = - {d^2\over {dx^2}} + x^2 + {\gamma \over {x^2}}
\end{equation}
with the well known spectral solution. The eigenvalues read $E_n
= 4n + 2 + (1+4\gamma )^{1/2}$, where  $n\geq 0$ and $\gamma >
-{1\over 4}$, with  eigenfunctions of the form:
\begin{equation}
f_n(x) = x^{(2\alpha +1)/2} exp(-{x^2\over 2})\, L^{\alpha
}_n(x^2)
\end{equation}
$$\alpha = {1\over 2}(1+4\gamma )^{1/2}$$
$$L_n^{\alpha } (x^2) = \sum_{\nu =0}^n {{(n+\alpha )!}\over
{(n-\nu )! (\alpha + \nu )!}} {{(-x^2)^{\nu }}\over {\nu !}}\, .$$

As in Example 2, we deal with  a clear double degeneracy of the
ground state and of the whole eigenspace of the self-adjoint
operator $H$. The singularity at $x=0$ decouples $(-\infty ,0)$
from $(0,+\infty )$ so that $L^2(-\infty,0)$ and $L^2(0,+\infty )$
are  the invariant subspaces for  dynamics  generated by $H$. We
encounter again two  \it separate \rm quantum problems
(degenerate ground state), respectively  on $R^+$ and $R^-$.\\

In the above example  the impenetrable barriers  are located at
the points where a potential singularity enforces the zero
boundary conditions. In particular,  such conditions are
satisfied  by (generalized) ground states and this feature is
mathematically responsible for the appearance of impenetrable
barriers. Indeed, to that end we can follow a rough argument. Let
$\phi \in L^2_{loc}(R^1)$ i. e. we consider all functions which
are square integrable on all  compact sets in $R^1$. If there is
a closed set $N$ of Lebesgue measure  zero so that  (strictly
speaking we admit  distributions)  ${{d\phi }\over {dx}} \in
L^2_{loc}(R^1 \setminus N)$, then there is  a uniquely determined
Hamiltonian $H$ such that $\phi $ is its (generalized) ground
state.  If $\phi \cdot (x-x_0)^{-1/2}$ is bounded in a
neighbourhood of $x_0$, then there is an impenetrable barrier at
$x_0$. For a precise description of this phenomenon in $R^n$, see
e.g. Ref. \cite{karw}.

 {\bf Example 4:} In contrast to the previous case where
the singularity of the potential alone was capable to make
 the ground state degenerate, due to the impenetrable barrier
at the origin, we can impose the existence of barriers as an
external boundary condition.
Let us introduce a differential expression
 $H_0= - {d^2 \over
{dx^2}}$ and observe that for any real $q$, the function $\psi
(x)= sin(qx)$ solves an equation $H_0\psi = q^2\psi $. The
operator $H_q= H_0 - q^2 $  is self-adjoint  when operating on
$D(H_q)= [f\in AC^2(R^1); f, f', f'' \in L^2(R^1), f({{n\pi
}\over q})=0, n=0, \pm 1, \pm 2,...]$  and $sin(qx)$ is its
generalized ground state. In that case a particle localized at
time $0$ in a concrete segment $((n-1){\pi \over q}, n{\pi \over
q})$ will be confined there forever.  This model can be
considered as that of multitrapping enclosures, with impenetrable
barriers at points $n{\pi \over q}$.
\\

There is one distinctive feature shared by the above exemplary
models: the Hamiltonian is a well defined self-adjoint operator in
each case,  respecting various confinement  (localization) demands.
There is however \it no \rm self-adjoint "momentum"-looking
 operator that would be compatible  with the trapping
 boundary conditions and the
corresponding  unitary time evolution rule in a trap.

\section{Quantum "life" in $L^2([0,\pi ])$}

Presently we shall devote more attention to self-adjoint operators  which can
be associated with differential expressions  $-{{d^2}\over {dx^2}}$ and
$-i{{d}\over {dx}}$ in $L^2([0,\pi ])$. We shall also  spend a while on  an
issue of their physical interpretation. That derives from the fact that there exists
a well developed  mathematics for various operators localized "on the interval",
while their physical relevance  is a matter of a specific context:  different
boundary data refer to an entirely  different physics.

Let us reconsider the Hamiltonian versus momentum operators interplay in  $L^2(R^1)$.
The standard differential expressions,  when acting on the space
$C_0^{\infty }(R^1)$ of
the infinitely differentiable functions of compact support, define symmetric operators.
Since $C_0^{\infty }(R^1)$ is invariant under differentiation, the operator
$-{{d^2}\over {dx^2}}$ can be interpreted as the "square" of $-i{{d}\over {dx}}$,
in the sense of its two consecutive actions.

In order to obtain self-adjoint operators we have a priori two possibilities:\\

(i) We extend  the symmetric operator $-i{{d}\over {dx}}$ to a self-adjoint operator
$\hat{p}$ which may be called a momentum operator,  and \it then \rm define the Hamilton
operator $H= {\hat{p}}^2$  where  the square is taken  in the sense of the spectral theorem. \\

(ii) Extend  the symmetric operator  $-{{d^2}\over {dx^2}}$ to a self-adjoint operator $H'$
which may  be called  a Hamilton operator, and \it then \rm  define the momentum operator
$\hat{p}' = (H')^{1/2}$ where the square root is taken in the sense of the spectral theorem.\\

As is well known  these two procedures give the same results:
$H=H',\, \hat{p}= \hat{p}'$ if considered in $L^2(R^1)$.

The situation appears to be different, when we pass to
$L^2([0,\pi ])$. The differential expressions when  acting in
$C_0^{\infty }(0,\pi )$ (now we restrict  the support to be
included in the  open interval $(0,\pi ) \subset R^1$) define
symmetric operators in $L^2([0,\pi ])$. Obviously, $C_0^{\infty
}(0,\pi )$ is invariant under differentiation and both procedures
(i) and (ii) can be safely utilized, except for the fact that
their outcomes (self-adjoint operators) \it no longer \rm
coincide.

In what follows we shall refer to the Krein - von Neumann theory of self-adjoint
extensions.  Let us begin from the case (i).

The closure of $-i{{d}\over {dx}}$ as defined on $C_0^{\infty }(0,\pi )$  is a closed
symmetric operator $\overline{p} = -i{{d}\over {dx}}$ with the domain $D(\overline{p}) =
\{ \psi \in AC[0,\pi ]; \psi (0)=0=\psi (\pi )\}$.

The deficiency index of $\overline{p}$ is $(1,1)$ and thus it has a one parameter family
of self-adjont extensions:
\be
p_{\alpha } = -i{{d}\over {dx}}
\ee
$$
D(p_{\alpha }) = \{ \psi \in  AC[0,\pi ]; \psi (0)= \exp(i\alpha ) \cdot \psi (\pi ) \}
$$
$$
0 \leq \alpha < 2\pi \, .
$$

For each chosen $\alpha $ there is in $L^2([0,\pi ])$ an orthonormal  basis which
is  composed of eigenvectors of $p_{\alpha }$:
\be
e^{\alpha }_n(x) = {1\over \sqrt{\pi }} \exp i(2n+{\alpha \over \pi })x
\ee

where $n$ takes integer values, while  the eigenvalues of $p_{\alpha }$ read:

\be
p_n^{\alpha } = 2n + {{\alpha }\over {\pi }} \, .
\ee

That allows to introduce another definition of $D(p_{\alpha })$.
Namely, if $f\in L^2([0,\pi ])$ is expressed in terms of
$e^{\alpha }_n$: \be f(x) = \sum_n f^{\alpha }_n e^{\alpha }_n(x)
\ee then  $f\in D(p^{\alpha })$ if an only if \be \sum_n
n^2 |f^{\alpha }_n|^2   < \infty \, . \ee

Now, $H_{\alpha }$   defined by
\be
H_{\alpha } = (p^{\alpha })^2 \, ,
\ee
in the sense of the spectral theorem, has the same family of eigenvectors as
$p^{\alpha }$,  but its eigenvalues read
\be
E^{\alpha }_n = (p^{\alpha }_n)^2 = (2n + {\alpha \over \pi })^2
\ee
for all integer $n$. (We recall that in the infinite well case we would have
$E_n \sim (n+1)^2$ where $n$ is a  natural  number.)

As a consequence,
\be
D(H_{\alpha })  = \{ f = \sum_n f^{\alpha }_n e^{\alpha }_n ; \sum_n
n^4 |f^{\alpha }_n|^2  < \infty  \}
\ee
and $D(H_{\alpha }) \subset D(p^{\alpha })$ and $D(p^{\alpha }) =
p^{\alpha } D(H_{\alpha })$.  Therefore the operator $H_{\alpha }$, Eq. (10) can be
safely interpreted as two consecutive actions of $p^{\alpha }$, Eq. (5) where
both operators are self-adjoint.
Also, there follows that
\be
H_{\alpha } =  -{{d^2}\over {dx^2}}
\ee
$$
D(H_{\alpha }) = \{ f\in AC^2[0,\pi ]; f(0)= \exp (i\alpha ) \cdot f(\pi ), f'(0) =
\exp(i\alpha ) \cdot f'(\pi )\} \, .
$$

Notice that in the special case of $\alpha =0$ one ends up with a  degenerate
spectrum $E_n = (2n)^2$,  where $n$ takes integer values. That corresponds to
the familiar plane rotator problem.

Now we turn to the procedure (ii).\\
The closure of  $-{{d^2}\over {dx^2}}$ as defined on $C_0^{\infty }(0,\pi )$ is
$\overline{H} = -{{d^2}\over {dx^2}}$, $D(\overline{H} = \{ \psi \in AC^2[0,\pi ]; \psi (0) =
\psi (\pi )= \psi '(0)= \psi '(\pi ) = 0\}$.
This is a closed  symmetric operator with the defect index $(2,2)$. Thus, the family of all
self-adjoint extensions of $\overline{H}$ is in one-to-one correspondence with $U(2)$,
the family of all $2\times 2$ unitary matrices.

To elucidate this correspondence, let us denote by $N_+$ the two-dimensional subspace of
$L^2([0,\pi ])$ with the orthonormal basis:
\be
\psi ^1_+(x) = (e^{2\pi } - 1)^{-1/2} \exp [(1-i)x]
\ee
$$
\psi _+^2(x) = (1- e^{-2\pi })^{-1/2} \exp [-(1-i)x]
$$
and analogously, we set $N_-$ for the linear span of:
\be
\psi _-^1(x) = (e^{2\pi } - 1)^{-1/2} \exp [(1+i)x]
\ee
$$
\psi _-^2(x) = (1-e^{-2\pi })^{-1/2} \exp [-(1+i)x]\, .
$$

Now define the map $I: N_- \rightarrow N_+$:
\be
I\psi ^{1,2}_- = \psi ^{1,2}_+ \, .
\ee

Given $U \in U(2)$, then $W = U \cdot I: N_- \rightarrow N_+ $ is unitary.
The self-adjoint extension of $\overline{H}$ corresponding to $U$ is defined by:
\be
D(H_U) = \{ g = f + w_- + Ww_- ;\,  f\in D(\overline{H}), w_- \in N_-\}
\ee
$$
H_U g = \overline{H} f - 2iw_- + 2iWw_- \, .
$$

In particular, by setting $U = - 1 $ where $1$ stands for the
unit  $2\times 2$ matrix, we obtain $\psi ^{1,2}_-(x) + W\psi
^{1,2}_-(x) = \psi ^{1,2}_-(x) - \psi ^{1,2}_+(x)$ which in view
of  Eqs. (14), (15)  yields $\psi^{1,2}_-(0) + W \psi^{1,2}_-(0) =
\psi ^{1,2}_-(\pi ) + W\psi ^{1,2}_-(\pi ) =0$. Clearly, $f \in
D(\overline{H})$ implies $f(0) =  0 = f(\pi )$. Accordingly, the
choice of $U = -1$ is equivalent to the infinite well boundary
conditions and thus we can specify the corresponding infinite
well Hamiltonian as follows:
\be D(H_{-1}) = \{ g\in AC^2[0,\pi ];
g(0)=g(\pi ) = 0 \}
\ee
$$
(H_{-1}g)(x) = - {{d^2}\over {dx^2}} g(x)\, .
$$

Now let us  define the $2\times 2$ matrix $U_{\alpha }$ with matrix elements:
\be
(U_{\alpha })_{11} = (U_{\alpha })_{22} = - {{1+i}\over 2}
\ee
$$
(U_{\alpha })_{12}(\chi ) =
{{i-1}\over 2} \chi {{(1+\chi \exp(\pi ))}\over {(1+ \overline{\chi }\exp(\pi ))}}
$$
$$
(U_{\alpha })_{21}(\chi ) = (\overline{U}_{\alpha })_{12}(\chi )
$$
where $\chi  = \exp(i\alpha ) $, $0\leq \alpha < 2\pi $ and
$\overline{\chi }$ and $\overline{U}_{\alpha }$ stand for complex
conjugates of $\chi$ and $U_{\alpha }$ respectively.

By inspection we can verify  that this choice of $U_{\alpha }$ is
equivalent to the boundary conditions $g(0) = \exp(i\alpha )\cdot
g(\pi ), g'(0) = \exp(i\alpha ) g'(\pi )$  and thus defines
$H_{U_{\alpha }} = H_{\alpha }$, Eq. (13)  or equivalently Eq.
(10),  with the domain $D(H_{\alpha })$, Eq. (12).

There is clearly \it no \rm apparent physical interpretation for
$H_\alpha $ and $p_\alpha $ in the context of the  infinite well,
or more generally - impenetrable barriers context. Nevertheless,
there are physical circumstances  under which  those operators
appear quite naturally, like e.g. the Aharonov-Bohm effect and an
involved quantum mechanics on multiply connected configuration
spaces, \cite{carlen}. That refers e.g.  to a charged particle in
the vicinity of an infinite cylindrical (eventually infinitely
thin) solenoid, when the parameter $\alpha $ in $H_{\alpha }$ can
be directly related to the magnetic flux inside  the solenoid.

Other instances when operators analogous to $H_\alpha $ are
relevant, refer to periodic potential models where e.g. $V(x)
\rightarrow \sum_k V(x+ k\pi )$ and  $V(x)$ is a continuous
function with $supp\,  V \subset (0,\pi )$.
 With the  vanishing (zero)  boundary condition  at $\pm \infty $
 imposed on its domain, the corresponding Hamiltonian is
 a self-adjoint operator.
Mathematically rigorous treatment  of  the closely related
Kronig-Penney model ($V(x)$ is  replaced by $\delta (x)$)  can be
found in Ref. \cite{holden}.

Let us define in $L^2([0,\pi ])$  the following self-adjoint operator:
\be
H_{V,\alpha }= - {{d^2}\over {dx^2}} + V(x)
\ee
$$
D(H_{V,\alpha }) = \{ g_\alpha \in AC^2[0,\pi ]; g_\alpha (0) = \exp(i\alpha )
\cdot g_\alpha (\pi ), g'_\alpha (0) = g'_\alpha (\pi ), 0\leq
 \alpha < 2\pi \}\, .
$$

As can be readily shown the Hilbert space $L^2(R^1)$ can be
unitarily mapped onto a direct integral of copies of $L^2([0,\pi
])$ with integration extending over  the segment $[0,2\pi ]$. The
corresponding direct integral  of operators $H_{V,\alpha }$:
 \be
{1\over {2\pi }} \int_{[0,2\pi )}^\oplus [H_{V,\alpha }] d\alpha
\ee
is then equivalent to $H_V$. In particular, when $V\equiv 0$
then $H_{V,\alpha }=H_\alpha $ and the direct integral (21) is
equivalent to $- {{d^2}\over {dx^2}}$ in $L^2(R^1)$.

\section{Conclusions}

The foundations of quantum mechanics employ both   the precision
of modern mathematical language and an elusive albeit deep
intuition based on an analysis of physical phenomena. The
major developments in quantum  theory and its  ability of a
 successful description of the microworld  owe more to the physical
 intuition than  to a precision of the mathematical apparatus.
That may presumably stand for a convincing justification of the widespread
attitude  towards  the usage (or rather  neglect) of sophisticated mathematical
 arguments.
Although we can regard a correspondence between
observables and self-adjoint operators in the Hilbert space as
generally accepted, the care for a precise formulation of the operator
domains is often considered as  an unnecessary nuisance or mathematical
pedantry.

On the other hand, mathematically oriented physicists  argue that the domain
subtleties in the operator analysis do carry a crucial physical information
 and must not be disregarded.
There seems  to be no  efficient interplay in the literature between those
 two (diverging) options: intuitive and rigorous. That is exactly the reason
 of so many "clashes" and "paradoxes" identifiable even in most trivial
 quantum mechanical problems.

 In the context of impenetrable barriers, the canonical quantization issue
 needs to be under scrutiny.  That pertains mostly to operators in $L^2(a,b)$
 where the segment $(a,b)$ is bounded from at least one side.
In the canonical quantization scheme, the correspondence principle
$x\rightarrow \hat{x}, \, p \rightarrow \hat{p} = -i\hbar
{{d }\over {dx}}$
had been  originally introduced in $L^2(R^1)$.
Under those circumstances, the intuitive definition of $\hat{x}, \hat{p}$ on
smooth functions  with reasonable fall-off at infinity is sufficient  to
 determine them uniquely as self-adjoint operators which obey  the canonical
 commutation relations in the Weyl form.

 That statement is purely kinematical and thus independent of any dynamics.
 The fact that $\hat{p}$ commutes with the free Hamiltonian $- {{\hbar }^2\over {2m}}
 {{d}^2\over {dx}^2}$ and thus is a constant of motion for a free
 particle, clearly supports the view that $\hat{p}$ is the  momentum operator.

The Hilbert space $L^2(a,b)$ has \it no \rm a priori physical interpretation. Its
physical meaning is closely related to that information on the dynamics which
is encoded in the boundary conditions at $a$ and $b$.

Summarizing  our observations let us  invoke  most frequently discussed
cases (with their own plethora of "paradoxes").

(i) The boundary conditions $f(a)=0=f(b)$  correspond to  the infinite
well problem,
and/or to the particle restricted to stay  in a semibounded  segment.
In that case, $L^2(a,b)$ is a subspace of $L^2(R^1)$ which is left invariant
 by the corresponding dynamics. The momentum operator clearly \it is \rm
 a measurable quantity, but is defined in the  encompassing "mother" space
$L^2(R^1)$.
There is no self-adjoint momentum operator  in $L^2(a,b)$ that would correspond
to $-i\hbar {{d}\over {d x}}$ and was at the same time  compatible with
the above boundary conditions.

(ii)  The boundary conditions of the type (13) or (20) (up to suitable rescalings)
correspond to the dynamics on $S^1$. The self-adjoint operator defined by
$-i\hbar {{d}\over {d\phi }}$ and the periodic boundary conditions  ($\alpha =0$)
corresponds to the angular momentum operator of the plane rotator. In case of
$\alpha \neq 0$  we deal with  the rotational observable for  a particle
rotating freely around an infinitely thin solenoid. The parameter $\alpha $
value is related to a magnetic flux, \cite{carlen}.

(An analogous reasoning can be carried over to higher dimensions, for
quantum particles  constrained to remain on a certain manifold.
Plane billards are typical examples in this context.)

Perhaps the most important outcome of our discussion  is that,
 even in the simplest conceivable  models  of restricted (trapped particle)
quantum systems,   it is illegitimate  to view a particle in the  trap
as an isolated  \it small \rm  "mesoscopic quantum world" and ignore
the existence of  its  \it large \rm   complement (exterior).
(We ignore anyway
all of the Universe, importance or lack of importance
attributed to the external observer, classical-quantum interplay, decoherence
 and an infinity of related conceptual issues.)

Specifically, we make a sharp distinction between the primordial kinematic
observables, whose eigenvalues are identical with classical phase-space
 labels, and  the emergent energy observable, the Hamiltonian which may
involve most sophisticated restrictions in the form of specific boundary
data or  general constraints.
Classically or semiclassically, that is exactly the point where the
emergent (!)  phase-space structure/topology
(interval or $S^1$ in $R^1$, rectangle or  cylinder in $R^2$) would
intervene.

\end{document}